# Solid-State Inrush Current Limiter Controller Based on Inrush Prediction for Large Transformers

Nima Tashakor [1*], Teymoor Ghanbari [2], Ebrahim Farjah [2], Ehsan Bagheri [2], Stefan Götz [1]

[1] TU Kaiserslautern, Kaiserslautern, Germany
[2] Shiraz University, Shiraz, Iran
[*] tashakor@eit.uni-kl.de

*Abstract–* Inrush current problems of the industrial transformers are occasionally more serious than the utility transformers. A new Solid-State Inrush Current Limiter (SSICL) is proposed for medium voltage power transformers. The SSICL consists of three similar sets. Each set includes a bidirectional semiconductor switch and an inrush current limiter resistor. Using a control strategy based on Kalman filtering, transformer inrush current is suppressed during magnetization process. In fact, built up profile of the inrush current is dictated in accordance with the estimated current by a slow dynamic Kalman filter.

A single-phase prototype SSICL is developed and tested. The proposed control scheme is simulated in MATLAB-SIMULINK. By using Real Time (RT) toolbox of the MATLAB software and a PCI_1711U, interrelationship between the SSICL and its controlling system is realized. Some experiments are carried out to evaluate the performance of the SSICL. The results show that the proposed SSICL can considerably suppress inrush current of the transformer.

# I. Introduction

High inrush current of electrical machines and industrial transformers during energization is one of the most concerns in distribution networks. The inrush current is a high-magnitude, unbalanced, and harmonic-rich current with a considerable DC component [1, 2]. Major problems relevant to inrush current are inadvertent operation of the transformer protection relays, decrease of transformer lifetime, mechanical and thermal stresses, power quality degradation, and transient overvoltage [3]. In some cases, such as offshore industrial transformers, inrush current has more serious problems than the utility transformers. In such transformers, inrush current may lead to transient overvoltage due to excitation of the low resonant frequency introduced by the large shunt capacitance of the sub-sea cables [4]. In addition to magnetization of the transformers, capacitor bank switching, and heavy inductive loads startup are two other sources of the inrush current. Some similar inrush current limiter devices have been also proposed for these applications in recent years [5-7].

The magnetizing current depends on several factors, including remnant flux at the switching instant, switching inception angle, resistance of the primary circuit, magnetizing characteristics of the transformer core and so on [3]. Based on these factors, some remedies/provisions/methods have been recommended to suppress or eliminate the inrush current. In recent years, many investigations have accomplished to limit inrush current of the power transformers [8-18].

The proposed techniques can be classified into the methods based on improvement of the transformer structure and the methods based on utilizing supplementary devices and/or control circuits. Elimination of the residual flux using high quality core material and series compensator [8], asymmetrical winding configuration [9], virtual air gap [10], and increase the transient inductance of the primary coil by changing the distribution of the coil winding [11] are some of the structural based methods. Topology-based methods lead to more complexity and cost. Furthermore, most of them could not be employed in already fabricated transformers. Insertion of a resistor/inductor and controlled switching of the transformer are two well-known approaches based on using the supplementary and/or control circuits [8], [12-15]. Main drawbacks of the methods based on controlling the instant of switching include need for prior knowledge of accurate remnant flux values, difficulty relevant to instantaneous measurement of the transformer residual flux, and uncertainty factors in the circuit breaker [9].

Power electronic based supplementary devices for limiting inrush current are attractive and could provide some more flexibility than the above-mentioned methods [16-20]. Most of the proposed supplementary devices and/or control circuits techniques could be implemented by using power electronic switches to have more capabilities. In [16, 17] a diode-bridge type DC reactor inrush current limiter and in [18] a resonate-type Solid-State Inrush Current Limiter (SSICL) have been suggested. Fully controlled switches such as Isolated Gate Bipolar Transistor (IGBT) and Integrated Gate-Commutated Thyristor (IGCT) could open a new window of opportunity designing higher efficient inrush current limiters. The main challenge of the power electronic based inrush current limiters is inaccessibility of high rating switches. Fortunately, fully controllable switches are accessible for the typical ratings of the industrial transformers.

In this paper, a high performance SSICL with a simple structure is proposed to limit high level of inrush current. The proposed three-phase SSICL consists of three similar sets, in which the inrush current is limited by three bidirectional semiconductor switches in parallel with three inrush current limiter resistors. The switches are controlled by using an algorithm based on Kalman filtering. The switching template is dictated in accordance with the estimated current by a slow dynamic Kalman Filter (KF). A residual signal is defined as difference between the estimated signal and measured current. Since the estimated signal by the slow KF cannot track abrupt change of the measured current, the residual signal has considerable amplitude during energization of the transformers. During magnetization condition, the residual signal determines the switching template, while in normal condition the residual signal becomes zero and so the switches are closed continually.

The overall concept of the three-phase SSICL is discussed and a single-phase prototype SSICL is analyzed in detail. Experimental results are presented to confirm the capabilities of the proposed SSICL. The control strategy is simulated in MATLAB-SIMULINK and other components of the SSICL are implemented in hardware. By using Real Time (RT) toolbox of MATLAB software, interrelationship between the hardware and software sections of the SSICL is realized. In fact, by using a PCI_1711U, the SSICL is controlled in real-time.

The rest of the paper is organized as follows: in section II, operating principles of the single-phase structure SSICL are discussed and its three-phase structure is introduced. In section III, basis of the proposed switching strategy is presented. The performance of the prototype single-phase SSICL is evaluated by some experiments in section IV. In section V the main results are concluded.

## II. The SSICL structure and principle

The structure of the single-phase SSICL is shown in Fig. 1 (a). It consists of a fast-acting bidirectional switch realized by power electronic semiconductors such as IGBT or IGCT with anti-parallel diodes, an inrush current resistor, a varistor (nonlinear resistor), and a snubber circuit. In addition to the fully-controlled bidirectional switch shown in Fig. 1, there are some other alternative bidirectional switches. The selected structure is chosen based on some considerations such as the conduction and switching losses as well as transients relevant to the reverse recovery current of the diodes. The line current is measured by using a Current Transformer (CT) and it is estimated by a slow dynamic Kalman Filter (KF). To cancel out any noise, output of the CT is filtered by using a Low-Passed Filter (LPF) as shown in Fig. 1.

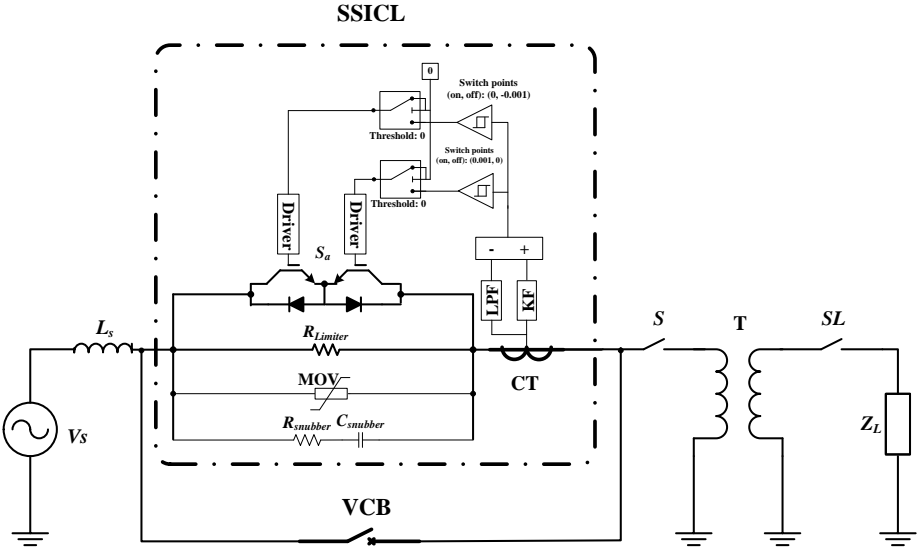

(a)

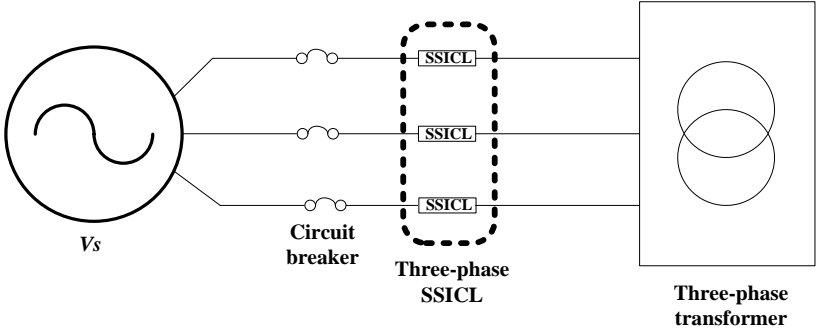

(b)

Fig.1. a) Single-phase structure of the SSICL b) Three-phase structure of the SSICL

The estimated signal is employed to reduce conduction time intervals of the IGBTs during energization of the transformer. A residual signal is generated by subtracting the estimated signal from the output signal of the LPF. When the residual signal exceeds a preset values ($\pm I_{th}$), the switches are turned off and the current is diverted to the limiter resistor, snubber circuit, and varistor. Consequently, the inrush current is suppressed by the limiter resistor, when the residual signal increases in energization condition. The snubber circuit and varistor are employed to suppress transient overvoltage of the switching. After several cycles, depending on the KF parameters, the residual signal is attenuated and therefore conduction time intervals of the IGBTs increases, whereas conduction time interval of the limiter resistor decreases. Finally, the bidirectional switch conducts the line current continually and bypasses the limiter resistor in steady state.

Since the IGBTs conduct permanently during normal condition, there is a conduction power loss about 1% of the rated load power in the switches. Therefore, the bidirectional switch is bypassed by using a Vacuum Circuit Breaker (VCB), in steady state. Then, the SSICL could be added in parallel with an already installed circuit breaker as shown in Fig. 1 (a). Three-phase structure of the SSICL consists of three similar sets as shown in Fig. 1 (b). The control circuit of the each set independently drives the related SSICL to suppress inrush current of the relevant line.

### III. The SSICL design considerations

*A. The SSICL power components designing*

A typical transformer current in energization condition is shown in Fig. 2. The transformer is switched on at $t_0$ and is charging during $t_0$-$t_2$. Core of the transformer at $t_1$ gets saturated and magnetizing inductance changes from $L_{NS}$ to $L_S$. During $t_1$-$t_2$ the inrush current begins to rise with a larger rate.

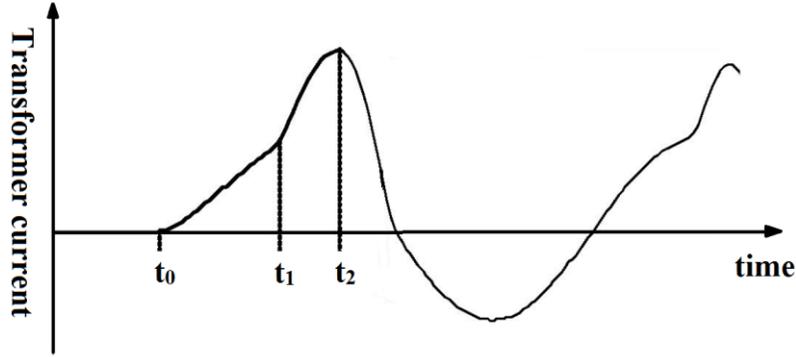

Fig. 2. Typical inrush current

Therefore, the residual signal exceeds ±$I_{th}$ during main portion of a cycle. In this condition, the IGBTs are gated on, but they cannot conduct. In fact, each triggered IGBT conducts when its anode-cathode voltage is negative. Thus, the line current passes through the limiter resistor. Gradually during the portion of a cycle at which the residual signal exceeds ±$I_{th}$, decreases (what decreases?). Consequently, the share of the line current conduction through the bidirectional switch increases. Therefore, two different modes are considered for the SSICL during inrush current limitation. a) Time interval at which the bidirectional switch is on b) Time interval at which the bidirectional switch is off. The line current equations in states (a) and (b) can be presented by (1) and (2), respectively.

$$i_{on}(t) = [-\frac{V_m}{Z}\sin(\omega t_0 - \phi)]e^{-\frac{R}{L}(t-t_0)} + \frac{V_m}{Z}\sin(\omega t - \phi) \tag{1}$$

$$i_{off}(t) = [I_1 - \frac{V_m}{Z}\sin(\omega t_2 - \phi)]e^{-\frac{R}{L}(t-t_2)} + \frac{V_m}{Z}\sin(\omega t - \phi) \tag{2}$$

where, $Z = \sqrt{R^2 + (\omega L)^2}$, $\phi = \tan^{-1}(\frac{\omega L}{R})$, $i_{on}(t_0)=0$, and $I_1=i_{on}(t_2)$

In the above equations, $R$ and $L$ are the total series resistance and inductance in the circuit. Also, $t_0$ and $t_2$ denote the instants, at which the bidirectional switch is gated on and off, respectively.

The resistance $R$, in which $R_{Limiter}$ is predominant, is very important for limiting the inrush current. By a suitable value of $R_{Limiter}$, the inrush current could be dramatically suppressed and the switching transient overvoltage of the SSICL in the limiting operation mode can also be alleviated.

Considering $i_2$ as the maximum permitted inrush current (see Fig. 2):

$$R = \frac{L}{kT} Ln \frac{V_m}{Zi_2 - V_m} \tag{3}$$

where, $t_2-t_1=kT$ and $k$ is fraction of the power cycle $T$.

The proposed SSICL can be designed for voltage levels, at which fully-controlled semiconductors are accessible [20]. Considering availability of the commercial IGBT in the medium voltage level (2-36 kV), it seems that the proposed SSICL could be fabricated for this level. The other components of the BFCL are easily accessible in the medium voltage level. High rating semiconductor switches are commercially available with current rating up to 24 kA and voltage rating up to 4 kV. Furthermore, at medium voltage level the semiconductor switches need to suitable snubber circuit for their protection, which is not necessary for low voltage level. The varistor (MOV) and series resistor-capacitor snubber ($R_s$ and $C_s$) protect the bidirectional switch against transient overvoltage switching [21]. Operation voltage of the MOV is selected below the maximum permissible voltage of the IGBTs.

### B. Kalman filtering

Control of the proposed SSICL is based on inrush current estimation using a KF. By using a slow dynamic KF, the fundamental component of the measured current is estimated in real-time. A mathematical model of the current signal in state-space form is utilized as KF model. A discrete sinusoid modal signal with an arbitrary amplitude and phase is considered for the model. First, the measured current signal is considered as:

$$S_n = a\cos(\omega_0 n + \varphi) \tag{4}$$

where, $\omega_0 = 2\pi f_0/f_s$, in which $f_0$ and $f_s$ are the measured signal and sampling frequencies respectively and $n$ is the time index.

Using basic trigonometric identities, $S_n$ can be recursively computed as follows:

$$S_{n+1} + S_{n-1} = 2\cos(\omega_0)S_n + \psi_n \tag{5}$$

where, $\psi_n$ denotes an additive zero-mean random term, considered to represent the possible model errors, including slight amplitude, phase, or frequency deviations.

Current signal noise due to the measurement and other factors can be superposed on the pure modal signal, therefore:

$$y_n = S_n + v_n \tag{6}$$

where, $S_n$ is the fundamental component of the measured signal and $v_n$ is a zero-mean random term, representing the noises.

In order to construct a KF to estimate the fundamental component of the modal signal, the dynamic equation in (5) needs to be converted into state-space form. Therefore, (5) and (6) for the fundamental component can be rewritten as follows:

$$\begin{aligned} X_{n+1} &= MX_n + b\psi_n \\ y_n &= h^T X_n + v_n \end{aligned} \tag{7}$$

where, $X_n \doteq [x_n \ x_{n-1}]^{-1}$, $M \doteq \begin{bmatrix} 2\cos(\omega_0) & -1 \\ 1 & 0 \end{bmatrix}$, $b \doteq [1\ 0]^T$, $h \doteq [1\ 0]^T$, $T_s = f_s^{-1}$ and $K$ depends on $X/R$ ratio of the system.

The dynamical model is now completed. The fundamental component of the measured current signal could be estimated using classical iterative KF equations presented in the Appendix. The residual signal is calculated as:

$$R_n = S_{f_n} - y_n \tag{8}$$

where, $S_f$ is output of the LPF.

Consider stationary process models ($q_n = q$) and stationary observations ($r_n = r$) in steady state condition of the filter (see Appendix). Noise covariance matrixes $q$ and $r$ act as tunable parameters to balance the dynamic response of the filter. Ratio $q/r$ rather than their actual values determines the dynamic response. It can tune a balance between sensitivity to noise and dynamic response of the filter. Therefore, to have a slow KF, $q/r$ should be chosen a large value.

## IV. Experimental Results

To verify the performance of the proposed SSICL, the simple system shown in Fig. 1 is simulated and implemented in the lab. The system and prototype single-phase SSICL parameters are tabulated in Table I. In Table II and Fig. 3, parameters of the transformer $T$ and its V-I curve are presented, respectively. The transformer inrush current waveforms are derived by measurement and recording in MATLAB-SIMULINK with and without the SSICL. Through heuristic methods, the required parameter values of the system components are extracted and utilized in the experiments. In Fig. 1, inrush current condition occurs by closing switch S, when SL is open.

The SSICL is controlled using a PCI_1711U. In fact, the control circuit is simulated in MATLAB-SIMULINK and by using RT toolbox; the SSICL is controlled in real-time. The laboratory set-up, which is considered for testing the SSICL is shown in Fig. 4.

Switching of the transformer at zero crossings of the voltage waveform may lead to highest inrush currents. Furthermore, residual flux has a considerable impact on inrush current and determines the first peak of inrush current. Therefore, the capability of the SSICL in the zero crossing is studied for several different values of residual fluxes. A heuristic approach is adopted for assigning the same residual flux value in magnetization condition with and without the SSICL.

**Table I:** Prototype SSICL and the system parameters

| Symbol | Quantity |
|---|---|
| $V_s$ | 220 V |
| $L_s$ | 5 mH |
| $R_s$ | 1 Ω |
| $R_{Limiter}$ | 10 Ω |
| $C_{snubber}$ | 47 nf |
| $R_{snubber}$ | 15 Ω |
| $T_1$ | 220/380 |
| IGBT | BUP314D |
| MOV | 10N180M |

**Table II:** The transformer parameters

| Symbol | Quantity |
|---|---|
| $S_n$ | 2.3 KVA |
| $V_1$ | 220 V |
| $V_2$ | 380 V |
| Z | 1.13+2.2 Ω |
| $X_m$ | 169 Ω |
| $R_c$ | 2.29 KΩ |

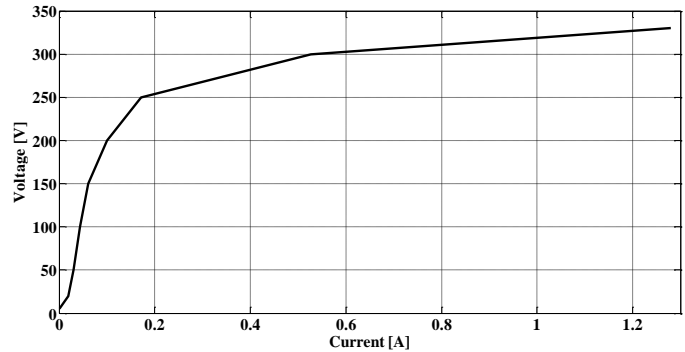

Fig.3. V-I curve of the transformer

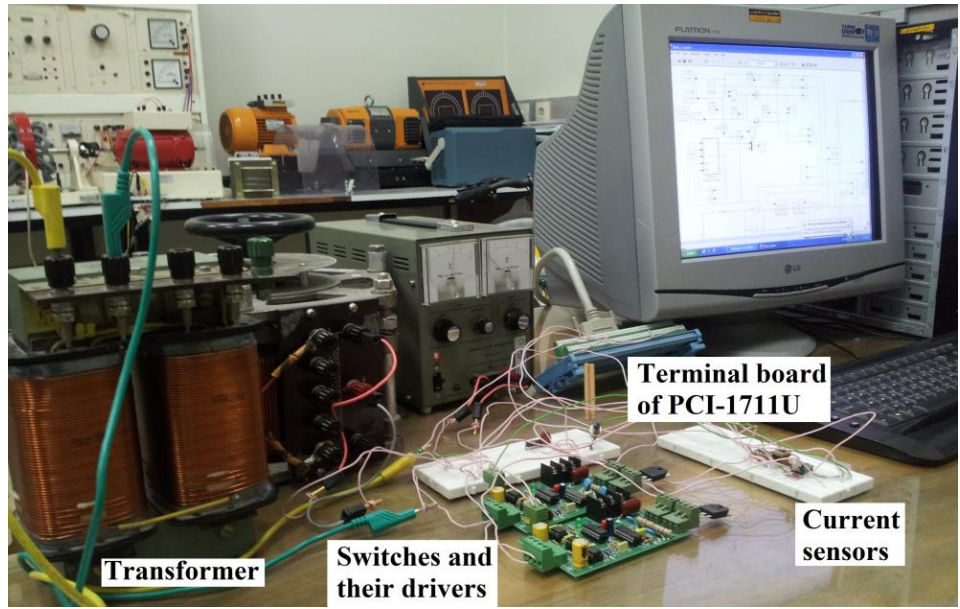

Fig.4. Test bench of a laboratory prototype of the SSICL

Fig. 5 shows a typical experimental result of the inrush current, when the transformer is switched on without the SSICL at zero crossing of the supply voltage. In Fig. 6, an inrush current with the SSICL at zero crossing of the supply voltage is presented. The inrush currents are recorded in several experiments with different residual flux values with and without the SSICL. In order to evaluate the performance of the proposed SSICL, one should provide the condition of maximum inrush current with and without SSICL. It should be noted that the remnant flux prior to start-up of the transformer is not known in practice, whereas the accurate remnant flux is needed for getting a fair evaluation of the SSICL. So, the only way to meet the above conditions is to repeat the energization process with and without SSICL for plenty of cases. Also, for each case by using a trial and error procedure, one should determine the appropriate remnant flux values. After repeating the above procedure for at least 200 times, it was concluded that inrush current with SSICL is almost limited to one-tenth of its value without SSICL.

Regarding Fig. 5 and Fig. 6, it can be observed that without the SSICL, the first peak of the current is about 25 A, which is limited to 2.5 A by the SSICL. In Fig. 7, the limited inrush current relevant to Fig. 6 and its estimated signal are presented. As shown in this figure, the first peak of the estimated signal is 1 A. The switching template dictates such that the inrush current is formed like this signal. After 0.04 sec, the estimated signal tracks the measured current accurately and the bidirectional switch closes continually. In Fig. 8, current of the bidirectional switch and limiter resistor relevant to limited inrush current of Fig. 6 are shown. Furthermore, switching template of the bidirectional switch in this case is shown in Fig. 9. This template is formed based on the generated residual signal relevant to limited inrush current of Fig. 6.

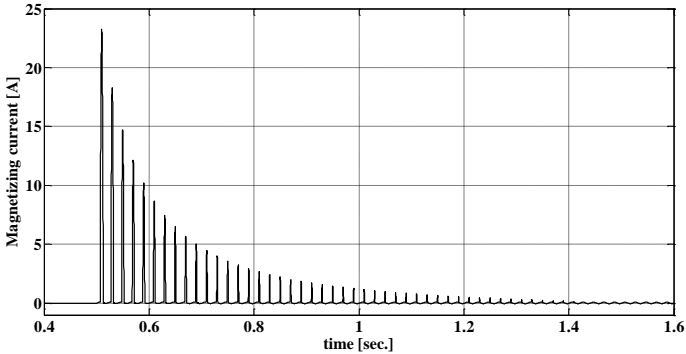
Fig.5. Inrush current without the SSICL

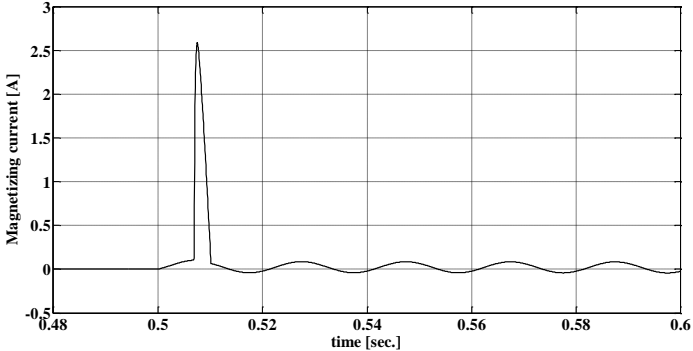
Fig.6. Inrush current with the SSICL

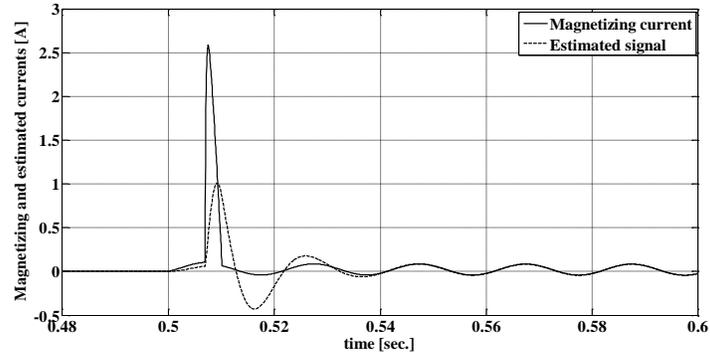
Fig.7. Inrush current and its estimated signal

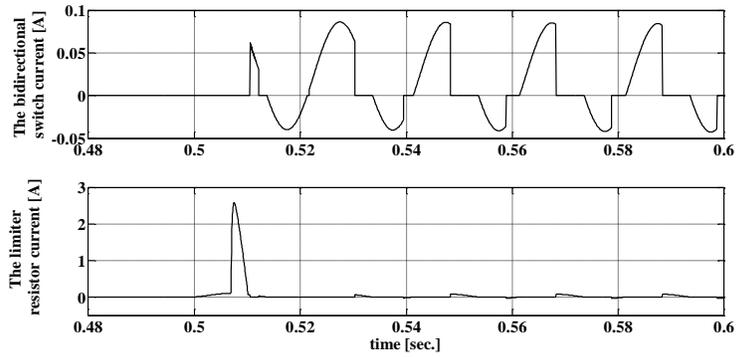
Fig.8. The bidirectional switch and limiter resistor currents

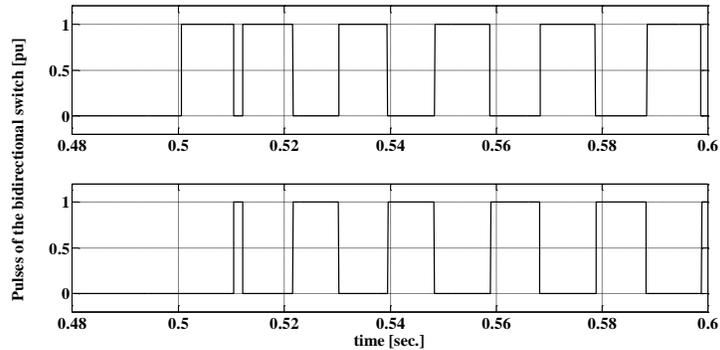
Fig.9. Pulses of the bidirectional switch

## V. Conclusion

A high performance SSICL is proposed in this paper. It is shown that by using a fully-controlled static switch in parallel with a limiter resistor, inrush current could be limited dramatically. The switch is controlled by using a novel algorithm based on a slow dynamic KF. The prototype

SSICL is capable of limiting the inrush current up to one-tenth of its value without SSICL. The main advantages of the proposed SSICL are low conduction loss in normal condition, fast operation, and simple structure. The experimental results reveal the performance and pragmatic features of the proposed SSICL.

## VI. Appendix

*Kalman filter process:*

1. Set initial estimate of the state vector and its error covariance matrix ($\hat{X}_n^-$ and $P_n^-$)

2. Compute the filter gain at instant *n* by: $K_n = \dfrac{P_n^- h}{h^T P_n^- h + r_n}$

3. Update the estimate using the measurement at instant *n* by: $\hat{X}_n^+ = \hat{X}_n^- + K_n(y_n - h^T \hat{X}_n^-)$

where, $q \doteq E\{\psi_n^2\}$, $r_n \doteq E\{v_n^2\}$, $\hat{X}_n^- \doteq \hat{E}\{X_n \mid y_{n-1}, ..., y_n\}$ is the *a priori* estimate of the state vector $X_n$ in the $n^{th}$ stage using the observations $y_1$ to $y_{n-1}$, and $\hat{X}_n^+ \doteq \hat{E}\{X_n \mid y_n, ..., y_1\}$ is the *a posteriori* estimate of this state vector after using the $n^{th}$ observation $y_n$. $P_n^-$ and $P_n^+$ are the state vector covariance matrices before and after using the $n^{th}$ observation, respectively and are defined similarly.

4. Compute the error covariance for the update estimate using: $P_n^+ = P_n^- - K_n h^T P_n^-$

5. Project the filter ahead as: $P_n^+ = P_n^- - K_n h^T P_n^- \hat{X}_{n+1} = A_n \hat{X}_n$, $\hat{P}_{n+1} = A_n P_n^+ A_n^T + q_n b b^T$

6. Start from 2.

# References

1. IEEE Std. C37.91, IEEE Power Eng. Soc.: 'IEEE Guide for Protecting Power Transformers', May 2008

2. M. J. Heathcote, 'The J&P Transformer Book', Elsevier, Oxford, U.K., 2007

3. Cheng, C.K., Chen, J.F., Liang, T.J., Chen, S.D.: 'Transformer design with consideration of restrained inrush current', *Electrical Power and Energy Systems*, 2006, 28, pp. 102-108

4. Smith, K.S., Ran, L., and Leyman, B.: 'Analysis of transformer inrush transients in offshore electrical systems', *IEE Proc.-Gener. Transm. Distrib.*, Jan 1999, 146, (1), pp 89-95

5. Ghanbari, T., and Farjah, E.: 'Development of an efficient solid-state fault current limiter for microgrid', *IEEE Trans. Power Del.*, 2012, 27, (4), pp. 1829-1834.

6. Ghanbari, T., Farjah, E., and Zandnia, A.: 'Development of a high-performance bridge-type fault current limiter', *IET Gener. Transm. Distrib.,* 2013, doi: 10.1049/iet-gtd.2013.0276, pp. 1–9

7. Ghanbari, T., Farjah, E., and Zandnia, A.: 'Solid-state transient limiter for capacitor bank switching transients', *IET Gener. Transm. Distrib.,* 2013, 7, (11), pp. 1272 – 1277

8. Shyu, J.L.: 'A novel control strategy to reduce transformer inrush currents by series compensator', Int. Conf. on Power Electronics and Drives Systems, PEDS 2005, vol. 2, pp. 1283–1288

9. Chen, J.F., Liang, T.J., Cheng, C.K., Chen, S.D., Lin, R.L., and Yang, W.H.: 'Asymmetrical winding configuration to reduce inrush current with appropriate short-circuit current in transformer', *IEE Proc. Electr. Power Appl.*, 2005, 152, pp. 605–611

10. Molcrette, V., Kotny, J.L., Swan, J.P., and Brudny, J.F.: 'Reduction of inrush current in single-phase transformer using virtual air gap technique', *IEEE Trans. Magn.*, 1998, 34, (4), pp. 1192–1194

11. Cheng, C.K., Liang, T.J., Chen, J.F., Chen, S.D., and Yang, W.H.: 'Novel approach to reducing the inrush current of a power transformer', *IEE Proc. Electr. Power Appl.*, 2004, 151, (3), pp. 289–295

12. Brunke, J.H., and Frohlich, K.J.: 'Elimination of transformer inrush currents by controlled switching, part I: Theoretical considerations', *IEEE Trans. Power Deliv.*, 2001, 16, (2), pp. 276–280

13. Brunke, J.H., and Frohlich, K.J.: 'Elimination of transformer inrush currents by controlled switching, part II: Application and performance considerations', *IEEE Trans. Power Deliv.*, 2001, 16, (2), pp. 281–285 14      Mahgoub, O.A.: 'Microcontroller-based switch for three-phase minimization'. Proceedings of the IEEE International power electronics congress, Cuernavaca, Mexico, 1996, pp. 107–112

15      Abdulsalam, S.G., and Wilsun X.: 'A Sequential Phase Energization Method for Transformer Inrush Current Reduction—Transient Performance and Practical Considerations'. *IEEE Trans. On Power Delivery*, Jan 2007, 22, (1), pp 208-216

16      Tarafdar Hagh, M., and Abapour, M.: 'DC reactor type transformer inrush current limiter', *IEE Proc. Electr. Power Appl.*, 2007, 1, (5), pp. 808–814

17      Madani, S.M., Rostami, M., Gharehpetian, G.B., and Haghmaram, R.: 'Inrush current limiter based on three-phase diode bridge for Y-yg transformers', *IEE Proc. Electr. Power Appl.*, 2012, 6, (6), pp. 345–352

18      Ghanbari, T., and Farjah, E.: 'Efficient resonant-type transformer inrush current limiter', *IEE Proc. Electr. Power Appl.*, 2012, 6, (7), pp. 429–436

19      N. Tashakor, B. Arabsalmanabadi, T. Ghanbari, E. Farjah and S. Götz, "Start-up circuit for an electric vehicle fast charger using SSICL technique and a slow estimator," in *IET Generation, Transmission & Distribution*, vol. 14, no. 12, pp. 2247-2255, 19 6 2020, doi: 10.1049/iet-gtd.2019.1477.

20      N. Tashakor, B. Arabsalmanabadi, F. Iraji and K. Al-Haddad, "Power sharing strategy for multi-source electrical auxiliary power unit with bi-directional interaction capability," in *IET Power Electronics*, vol. 13, no. 8, pp. 1554-1564, 17 6 2020, doi: 10.1049/iet-pel.2019.0715.

21      T. Ghanbari, E. Farjah and N. Tashakor, "Thyristor based bridge-type fault current limiter for fault current limiting capability enhancement," in IET Generation, Transmission & Distribution, vol. 10, no. 9, pp. 2202-2215, 9 6 2016, doi: 10.1049/iet-gtd.2015.1364.